\documentclass[prb,twocolumn,showpacs,amsmath,amssymb,aps]{revtex4}

\usepackage{graphicx}
\usepackage{dcolumn}
\usepackage{amsmath}
\usepackage{bm}

\begin{document}
\title{Enhanced Domain Wall Motion in the Spin Valve Nanowires}

\author{Tae-Suk Kim}
\affiliation{Department of Physics, Pohang University of Science and Technology, Pohang 790-784, Korea}
\received{\today}

\begin{abstract}
According to the recent experiment by the Fert group, the velocity of domain wall motion in the spin valve ferromagnetic nanowires 
was almost doubly enhanced compared to the old value. 
In this work, we propose an additional torque model, arising from the interlayer exchange interaction, which can enhance or 
suppress the domain wall velocity depending on the sign of the exchange constant or the wall motion direction relative to 
the magnetization orientation of the fixed layer. 

\end{abstract}

\pacs{75.45.+j, 75.25.-b, 85.75.-d}

\maketitle

 The dynamics of magnetic domain wall (DW) in the ferromagnetic nanowires 
has been a very active research area \cite{DW_motion} in spintronics. 
The DW motion under the magnetic field seems to be well documented experimentally and understood 
theoretically in terms of the phenomenological Landau-Lifshitz-Gilbert (LLG) equation.
After the Berger's pioneering work \cite{Berger1984} about the spin transfer torque, many experimental groups 
\cite{exp_DW1,exp_DW2,exp_DW3,exp_DW4,exp_DW5,exp_DW6} 
observed the magnetic DW motion under spin current and much theoretical efforts \cite{bazaliy,tatara,barnes,li_zhang,thiaville} 
have been exerted to refine the spin torque theory.
Though several theoretical issues are still unresolved, the experimental efforts are continued in order to 
enhance the DW velocity \cite{Parkin100, Fert180} and reduce the critical current density. \cite{exp_DW3}
These two factors are essential for the device applications in logic \cite{DW_logic} as well as in memory. \cite{DW_memory}

 Recently, a highly enhanced DW velocity under spin current was reported \cite{Fert180} in the spin valve ferromagnetic nanowires.
The spin valve nanowire consists of FeNi/Cu/Co/CoO layers. 
The Co layer is magnetized along the magnetic easy axis or the wire direction and its magnetization is uniform and fixed by exchange bias.
The magnetization in the NiFe layer is also directed into the wire direction and the DW is introduced.
The current flow in plane induces the DW motion. 
While the highest DW velocity reported \cite{Parkin100} till now was about 100 m/s in a ferromagnetic nanowire with one layer,
the observed DW velocity in the spin valve (with trilayer) \cite{Fert180} reached up to 180 m/s.
The spin valve nanowires seem to be very promising for the future device applications.

\begin{figure}[bp]
\includegraphics[width=8cm,angle=0]{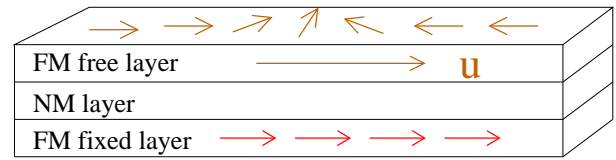}
\caption{(color online) Schematic structure of the spin valve nanowire. Two FM layers are separated from each other by the nonmagnetic (NM) layer. 
The magnetization in the bottom layer (fixed layer) is pinned along the wire direction, while the magnetization
in the top layer (free layer) is free to rotate. The magnetic easy axis is the wire direction. 
After introduction of DW in the free layer, DW is set in motion by spin current. 
The interlayer exchange coupling between two ferromagnetic layers can enhance or suppress the DW motion, 
depending on the sign of the exchange coupling and DW motion direction relative to the magnetization orientation of the fixed layer.}
\label{spinvalve}
\end{figure}

 In this paper we would like to propose another source of spin torque in the spin valve nanowires 
(Fig.~\ref{spinvalve}) which can enhance or suppress the DW velocity under spin current.
The two top (free) and bottom (fixed) layers are separated by the nonmagnetic spacer layer. 
In the magnetic multilayers, it is well known that there exists the interlayer exchange coupling.
Ever since the first observation \cite{Grunberg1986} of an antiferromagnetic coupling between two magnetic layers 
in the transition metal multilayers,
the interlayer exchange couplings as a function of the interlayer spacing have been observed \cite{osc_ilc} to be oscillating 
between ferromagnetic and antiferromagnetic couplings.
This interlayer exchange coupling is a source of an additional torque in the spin valve nanowires.
Though the theory of current-induced domain wall motion (CIDWM) is still very controversial, we adopt the generalized LLG equation under spin current 
in order to illustrate the enhanced or suppressed domain wall motion based on our model.

 The phenomenological LLG (Landau-Lifshitz-Gilbert) equation has been very useful in understanding the dynamics of 
magnetic domain walls under the magnetic field. This LLG equation was also used for studying the CIDWM by adding the spin transfer torque. 
In general, the LLG equation for magnetization is given by
\begin{eqnarray}
\frac{\partial}{\partial t} \vec{m} 
  &=& - \gamma \vec{m} \times \vec{H} + \alpha \vec{m} \times \frac{\partial}{\partial t} \vec{m}  + \vec{\tau}. 
\end{eqnarray}
Here $\vec{m}$ is the unit vector for magnetization of the free layer and $\vec{H}$ is the effective magnetic field (applied, exchange, and anisotropy fields).
The first term describes the precessional motion of $\vec{m}$ under $\vec{H}$, the second term is the Gilbert damping torque 
($\alpha$ is the dimensionless damping constant), and $\vec{\tau}$ is any additional torque acting on $\vec{m}$. 
Depending on the current flow direction, current in plane (CIP) or current perpendicular to plane (CPP), two models are proposed
\begin{eqnarray}
\vec{\tau}_{CIP} &=& - \vec{v}_s \cdot \vec{\nabla} \vec{m} + \beta \vec{m} \times (\vec{v}_s \cdot \vec{\nabla} ) \vec{m},  \\
\vec{\tau}_{CPP} 
  &=& - \gamma a_J \vec{m} \times (\vec{m} \times \vec{m}_s) - \gamma b_J \vec{m} \times \vec{m}_s.  
\end{eqnarray} 
In the CIP geometry, the first (second) term is (non)adiabatic torque and $v_s$ is proportional to the current density and the spin polarization. 
In the CPP geometry, $\gamma$ is the gyromagnetic ratio and both $a_J$ and $b_J$ are proportional to the current and 
the spin polarization, $\vec{m}_s$ is the magnetization unit vector in the fixed layer. 
The effect of $\vec{\tau}_{CPP}$ on the DW motion in spin valve nanowires was studied \cite{spinvalve_CPP} recently theoretically.
However. in the experiment \cite{Fert180} of the spin valve, the DW motion was measured in the CIP geometry, but not in the CPP geometry.
In this work, we study the LLG equation under the action of $\vec{\tau}_{CIP}$ and the interlayer exchange coupling. 
The interlayer exchange coupling is modeled by the following Hamiltonian \cite{interlayer}
\begin{eqnarray}
H_{layers} &=& - J \vec{m} \cdot \vec{m}_s.
\end{eqnarray}
For the positive (negative) exchange constant $J$, the coupling is ferromagnetic (antiferromagnetic). 
The sign of exchange coupling depends on the spacing of the nonmagnetic layer sandwiched in between two ferromagnetic layers. 
The above interlayer coupling gives rise to an additional torque on the magnetization $\vec{m}$.
\begin{eqnarray}
\vec{\tau}_{layer} &=& - \gamma J \vec{m} \times \vec{m}_s.
\end{eqnarray}
This torque is nonzero only near the domain wall and vanishes in the collinear sections of a nanowire.
This torque is very similar to the second term in $\vec{\tau}_{CPP}$, but there is a big difference: $J$ is nonzero even when there is no 
current perpendicular to the spin valve, while $b_J = 0$ without any perpendicular current.

 Our goal is to study the domain wall motion under the action of two torques $\vec{\tau}_{CIP}$ and $\vec{\tau}_{layer}$.
\begin{eqnarray}
\vec{\tau} &=&  - \vec{v}_s \cdot \vec{\nabla} \vec{m} + \beta \vec{m} \times (\vec{v}_s \cdot \vec{\nabla} ) \vec{m}
                         - \gamma J \vec{m} \times \vec{m}_s.
\end{eqnarray}
The effective magnetic field $\vec{H} = - \delta E/\delta \vec{m}$ is computed from the following magnetic energy density for magnetization 
\begin{eqnarray}
E &=& \frac{1}{2} Ja^2  (\partial_z \vec{m} )^2 - A m_z^2 + K m_y^2.
\end{eqnarray}
Here $J$ is the exchange coupling, $A$ and $K$ are the anisotropy constants, and $a$ is the lattice spacing. 
Representing the magnetization $\vec{m}$ in terms of two Euler angles $\theta$ and $\phi$ and minimizing the magnetic energy, 
a transverse wall is obtained. 
\begin{eqnarray}
\phi(z,t) &=& \Phi(t),  \\
\theta(z,t) &=& 2 \cot^{-1} \exp\left( - \frac{z-q(t)}{\Delta} \right).
\end{eqnarray}
Here $q$ is the DW position and $\Delta$ defines the domain wall width. 
\begin{eqnarray}
\Delta^2 &=& \frac{Ja^2}{ 2(A + K \sin^2 \Phi)}.
\end{eqnarray}
In order to see the effect of the interlayer exchange interaction, we consider the above rigid transverse domain wall.
Though the azimuthal angle depends on time, the time dependence of the domain wall width can be neglected in the leading approximation.
Finally the LLG equation in components is
\begin{eqnarray}
\frac{\dot{q}}{\Delta} - \alpha \dot{\Phi} 
  &=& \frac{v_s}{\Delta}  + \gamma K \sin2 \Phi,  \\
\alpha \frac{\dot{q}}{\Delta}  + \dot{\Phi} 
  &=& \frac{\beta v_s}{\Delta} + \gamma J.  
\end{eqnarray}
In the steady mode, $\dot{\Phi} = 0$ so that the domain wall velocity $u = \dot{q}$ is given by 
\begin{eqnarray}
u &=& \frac{\beta}{\alpha} v_s + \frac{\gamma J \Delta}{\alpha}.
\end{eqnarray}
The first term in $u$ comes from the spin transfer torque from spin current, while the second term from the interlayer exchange coupling.

 In the spin valve structure of a ferromagnetic nanowire, the domain wall motion is affected by the fixed layer
via the interlayer exchange interaction. the DW velocity can be either enhanced or suppressed depending on 
the relative direction of DW motion and the magnetization orientation of the fixed layer or the sign of the interlayer
exchange constant. 
For the ferromagnetic coupling, \cite{footnote}
the DW velocity is enhanced (suppressed) when its motion is directed along (against) the magnetization 
orientation of the fixed layer.
For the antiferromagnetic coupling, the DW velocity behaves in an opposite way compared to the ferromagnetic coupling case.

 Our result shows that the domain wall keeps moving even in the absence of driving current.
This is obvious because two magnetizations tend to be (anti)aligned for (anti)ferromagnetic coupling.
In real samples, the pinning potentials will tend to block the domain wall motion induced by the interlayer exchange.
The presence of the interlayer coupling will help enhance the DW motion above the critical current after overcoming the pinning potentials,
 if the condition is right.
In the experiment, \cite{Fert180} only one-way motion of DW (probably along the magnetization of Co layer)  under spin current is reported.
According to the interlayer exchange coupling, the wrong way motion of DW will be suppressed by the interlayer torque.
This will be a critical test of the interlayer exchange torque in the spin valve system.

 Our model study can be equally applied to the ferromagnetic nanowire with the magnetic tunnel junction structure.
Slonczewski \cite{mtj} predicted the exchange interaction between two ferromagnetic layers which are separated by the insulating barrier, 
even in the absence of the perpendicular current flow. 
Since the effect of the interlayer exchange coupling gets larger with larger couplings, the thinner insulating barrier will be more favorable.
Another strong point of the MTJ spin valves will be a weaker dependence of the interlayer exchange coupling on the current in plane.
In the case of nonmagnetic metallic spacer, the electric current may well generate the electron-hole excitations and disrupt 
the spin coherence so that the interlayer exchange coupling is reduced just like the thermal excitation.
At least, this disruption is absent for the insulating spacer.


\begin{thebibliography}{100}
\bibitem{footnote} In the text, the head-to-head DW is considered. For the tail-to-tail DW, the opposite results are obtained 
compared to the head-to-head DW.
\end{thebibliography}
\end{document}